\documentclass[journal]{IEEEtran}
\usepackage{amsmath,amsthm}
\usepackage[utf8]{inputenc} 
\usepackage[T1]{fontenc}    
\usepackage{url}            
\usepackage{booktabs}       
\usepackage{amsfonts}       
\usepackage{nicefrac}       
\usepackage{microtype}      
\usepackage{xcolor}         
\usepackage{graphicx}
\usepackage{caption}
\usepackage{cite}
\usepackage[colorlinks,bookmarks=false]{hyperref}
\hypersetup{citecolor=[RGB]{0,0,255}}

\newtheorem{proposition}{Proposition}
\newtheorem{definition}{Definition}
\newtheorem{theorem}{Theorem}
\newtheorem{corollary}{Corollary}

\newcommand{\dist}{\mathrm{dist}}

\usepackage{amsmath}
\usepackage{algorithmic}
\usepackage{array}
\begin{document}
\title{Stability Constrained Reinforcement Learning for Real-Time Voltage Control}

\author{Yuanyuan Shi$^{1,*}$, Guannan Qu$^{2,*}$, Steven Low$^{3}$, Anima Anandkumar$^3$ and Adam Wierman$^3$
\thanks{*Authors contributed equally.}
\thanks{$^{1}$Yuanyuan Shi is with the Department of Electrical and Computer Engineering, University of California San Diego, {\tt\small yyshi@eng.ucsd.edu}}%
\thanks{$^{2}$Guannan Qu is with the Department of Electrical Engineering, Carnegie Mellon University {\tt\small gqu@andrew.cmu.edu}}%
\thanks{$^{3}$Steven Low, Anima Anandkumar and Adam Wierman are with the Computing and Mathematical Sciences, California Institute of Technology.}%
}

\maketitle

\begin{abstract}
Deep reinforcement learning (RL) has been recognized as a promising tool to address the challenges in real-time control of power systems. However, its deployment in real-world power systems has been hindered by a lack of formal stability and safety guarantees. In this paper, we propose a stability constrained reinforcement learning method for real-time voltage control in distribution grids and we prove that the proposed approach provides a formal voltage stability guarantee. The key idea underlying our approach is an explicitly constructed Lyapunov function that certifies stability. We demonstrate the effectiveness of the approach in case studies,  where the proposed method can reduce the transient control cost by more than 30\% and shorten the response time by a third compared to a widely used linear policy, while always achieving voltage stability. In contrast, standard RL methods often fail to achieve voltage stability.
\end{abstract}


\IEEEpeerreviewmaketitle

\section{Introduction}
\label{sec:intro}

To achieve high penetration of distributed renewable generations at the distribution grid level, maintaining the voltage levels within the safe limit has been increasingly challenging.  Recently, significant efforts have been put into the design of real-time feedback controllers for voltage control purposes, see e.g. \cite{zhang2013local,bolognani2013distributed,li2014real,zhu2016fast,cavraro2016value,liu2018hybrid,tang2019fast,qu2019optimal,magnusson2020distributed}. 
Despite the progress, most of the existing work has only been able to optimize the steady state cost, i.e. the cost of the operating point after the voltage converges into the safe limit. In the meanwhile, transient performance is of equal importance. For example, once voltage violation happens, an important goal is to bring the voltage profile back to the safe limit as soon as possible, or in other words, to minimize the voltage recovery time. However,  optimizing or even analyzing the transient cost like the voltage recovery time has long been challenging as this is a nonlinear control problem. 
The challenge has further been complicated by the fact that many existing works \cite{li2014real,cavraro2016value,liu2018hybrid,tang2019fast,qu2019optimal,magnusson2020distributed} require knowledge of the topology and the parameters of the electric grid. Yet, such knowledge is not always available due to frequent system reconfigurations (resulting in topology change~\cite{weng2016distributed}) and time-varying system parameters resulting from facility aging, temperature and humidity change, etc. 

Reinforcement Learning (RL), having achieved impressive success in the past decade in game play~\cite{silver2016mastering}, robotics \cite{kalashnikov2018scalable}, is a promising tool to address the above challenges. RL methods do not need knowledge of explicit models and can learn from interactions with the underlying system. Further, due to the expressive power of neural networks as controllers/policies, RL is effective in learning nonlinear controllers with good transient performance. As a result, there has been tremendous interest in using RL for voltage control ~\cite{chen2020data,9353702,9328796,9143169,mukherjee2021scalable,kou2020safe,xu_optimal_2020,yang_two-timescale_2020,wang_data-driven_2020,wang_safe_2020,duan_deep-reinforcement-learning-based_2020,cao_multi-agent_2020,liu2020two}, see \cite{chen2021reinforcement} for a recent review. 


Despite this interest, it has been generally agreed that the key challenge in applying RL to power systems is the \emph{stability} issue \cite{chen2021reinforcement}. 
Specifically, power systems are critical infrastructure systems that place a high emphasis on stability, i.e. the ability to maintain at safe operating points under disturbances. Stability is important because instability can lead to unsafe operating conditions that violate regulatory requirements \cite{regulator} or even lead to catastrophic consequences, e.g., blackouts \cite{haes2019survey}. 
Despite the importance of stability, off-the-shelf RL algorithms lack provable stability guarantees. In particular, popular RL methods for continuous control, such as deep deterministic policy gradient (DDPG)~\cite{lillicrap2015continuous}, are gradient-based methods that focus on minimizing cost and do not explicitly consider stability. Even if the learned policy may appear ``stable'' on the training data set, it is not guaranteed to be stable as stability is a worst case concept requiring provably checking under the worst case scenario, which off-the-shell RL methods do not consider. 
The lack of provable stability guarantees is one of the biggest hurdles in applying RL to power systems since as mentioned earlier, instability can be catastrophic. 




Motivated by the challenge above, the question we address in this paper is: \emph{Can we apply RL to voltage control with provable stability guarantee?} 




\textbf{Contributions.} We answer the question affirmatively by designing a stability constrained RL framework 
that learns a control policy to optimize transient cost for voltage control with provable stability guarantees. The key idea underlying our approach is that we show strict monotonicity of the policy is sufficient to formally guarantee stability (Theorem~\ref{thm:voltage_stab}). The technique underlying Theorem~\ref{thm:voltage_stab} is that we derive an explicitly constructed Lyapunov function, which we use to certify stability for all monotone policies. With this stability result, we propose a Stable-DDPG approach which integrates the monotone constraint with DDPG through monotone policy network design (Corollary~\ref{corollary:stacked_relu}). 
The proposed method enables us to leverage the power of RL to improve the transient performance of voltage control without knowing the underlying model parameters, and in the meanwhile provably guarantee stability during and after the training. To the best of our knowledge, this is the first RL approach that learns nonlinear policies with stability guarantees for voltage control. 

We also perform numerical case studies to demonstrate the effectiveness and stability of the proposed method with both simulated disturbances and real-world data. Our method guarantees voltage stability under all operating conditions, which is not true for the standard DDPG method. In addition, our method can reduce the transient control cost by more than 30\% and shorten the voltage recovery time by a third compared to a widely used linear policy in the literature \cite{li2014real,zhang2013local}. 

\textbf{Related Work.} This paper connects to a broad set of literature in RL, control, and power systems.

\textit{Lyapunov-based Policy Learning.} The Lyapunov theory is a systematic framework to analyze the stability of a control system. The core idea is to identify a positive definite function (i.e., a Lyapunov function) of the system’s state, with negative derivatives along system trajectories~\cite{khalil2002nonlinear}. 
Using Lyapunov functions in RL was first introduced by~\cite{perkins2002lyapunov}, but the work did not discuss how to find a candidate Lyapunov function in general except for a case-by-case construction. A set of recent works including~\cite{chang2019neural,jin2020neural,chow2018lyapunov,richards2018lyapunov,manek2020learning} have attempted to address this challenge by jointly learning the policy and the Lyapunov function, where \cite{chow2018lyapunov} uses linear programming and \cite{chang2019neural,jin2020neural,richards2018lyapunov,manek2020learning} parameterizes the Lyapunov function as neural networks. 
In the context of these works, our contribution can be viewed as explicitly constructing a Lyapunov function for the voltage control problem and using it to guide policy learning. 



\textit{Reinforcement Learning for Power Systems.} 
Our work contributes to a growing line of papers that use RL for voltage control~\cite{chen2020data,9353702,9328796,9143169,mukherjee2021scalable,kou2020safe,xu_optimal_2020,yang_two-timescale_2020,wang_data-driven_2020,wang_safe_2020,duan_deep-reinforcement-learning-based_2020,cao_multi-agent_2020,liu2020two}, see \cite{chen2021reinforcement} for a recent review. As pointed out in  \cite{chen2021reinforcement}, one of the key issues in RL for power system control is the lack of provable stability guarantees, and our work makes a step toward addressing this issue by providing a formal stability guarantee on the learned policy. In particular, our experiments compare against~\cite{wang_data-driven_2020}, which uses standard multi-agent DDPG for voltage control. Closest in spirit to our paper is~\cite{cui2020reinforcement}, which proposes a stable RL approach for frequency control via a Lyapunov approach. However, their approach only applies to the frequency control application, while our method works for voltage control which requires a different Lyapunov function design. Interestingly, both our work and prior work~\cite{cui2020reinforcement} arrive at a similar stability condition, that is strict policy monotonicity guarantees system stability. 


\section{Preliminaries}
\label{sec:voltage}
In this section, we first introduce the distribution system power flow models and the voltage control problem formulation. Then, we review some background for policy optimization in reinforcement learning.
\vspace{-12pt}
\subsection{Branch Flow Model for Distribution Networks}
\label{sec:voltage_control_formuation}
\begin{figure}[ht]
    \centering
    \includegraphics[width=0.25\textwidth]{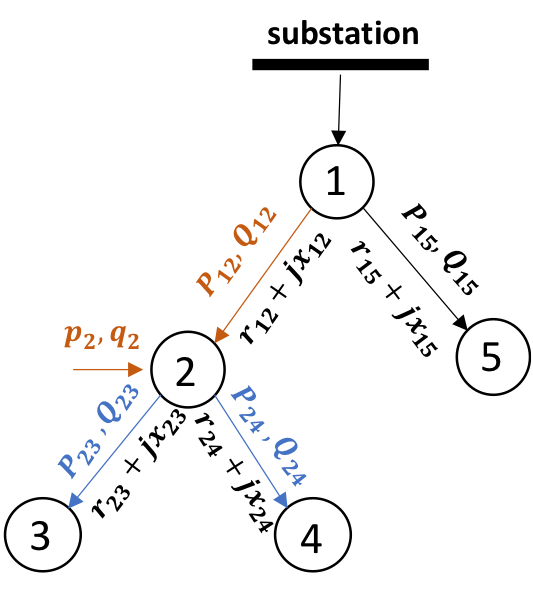}
    \caption{A 5-bus radial distribution network.}
    \label{fig1:vol_ctrl}
\end{figure}

We consider the distribution network as a tree-structured graph $\mathcal{G} = (\mathcal{N},\mathcal{E})$, which consists of a set of nodes $\mathcal{N}_0=\{0,1,\ldots,n\}$ and edges $\mathcal{E}$, where node $0$ is the substation. We use $\mathcal{N} = \mathcal{N}_0/\{0\}$ to denote the set of nodes excluding the substation node. See Fig. \ref{fig1:vol_ctrl} for an example 5-bus network.
Each node $i\in\mathcal{N}$ is associated with an active power injection $p_i$, a reactive power injection $q_i$, and a voltage magnitude $v_i$. We use $p, q$ and $v$ to denote the $p_i,q_i,v_i$ stacked into a vector. We consider the linear distribution power flow model, known as the Simplified Distflow model, which is a linear approximation of the underlying nonlinear branch flow model in radial distribution networks~\cite{baran1989network} by neglecting the quadratic power loss terms. The Simplified Distflow model is a commonly used model in the voltage control literature, see e.g.~\cite{li2014real,cavraro2016value,liu2018hybrid,tang2019fast,qu2019optimal,magnusson2020distributed} and the references therein. In the Simplified Distflow model, $p$, $q$ and $v$ satisfy the following equations, $\forall j \in\mathcal{N}, i=\textrm{parent}(j)$, \begin{subequations}
{\small
\begin{align}
    -p_j &= P_{ij}  - \sum_{k: (j, k) \in \mathcal{E}} P_{jk}\,, \quad
    -q_j = Q_{ij} - \sum_{k: (j, k) \in \mathcal{E}} Q_{jk}\,, \label{eq:conserv_law}\\
    -v_j &= v_i - 2(r_{ij}P_{ij} + x_{ij} Q_{ij})\,, (i, j) \in \mathcal{E} \label{eq:bfm_v}
\end{align}}
\end{subequations}
In Equation~\eqref{eq:conserv_law}, $P_{ij}$ and $Q_{ij}$ represent the active power and reactive power flow on line $(i,j)$, and $r_{ij}$ and $x_{ij}$ are the line resistance and reactance. Briefly speaking, Equation~\eqref{eq:conserv_law} presents the power conservation law at node $j$, e.g., in Figure~\ref{fig1:vol_ctrl}, the power inflow at node 2 (orange) equals to the power outflow at node 2 (blue). Equation \eqref{eq:bfm_v} models the voltage drop from node $i$ to node $j$. 
The above equation can also be rearranged and written in the vector form~\cite{li2014real}:
\begin{equation}
\mathbf{v} = R \mathbf{p} + X \mathbf{q} + \mathbf{v}_0 \mathbf{1} = X\mathbf{q} + \mathbf{v}^{env}. \label{eq:lindistflow}
\end{equation}
Here we separate the voltage magnitude $\mathbf{v}$ into two parts: the controllable part $X\mathbf{q}$ that can be adjusted via adjusting reactive power injection $\mathbf{q}$ through various control devices, and the non-controllable part $\mathbf{v}^{env} = R\mathbf{p}+\mathbf{v}_0$ that is decided by the load and PV power $\mathbf{p}$. Matrix $R, X$ are given as follows, $R_{ij} = 2 \sum_{(h, k) \in P_i \cap P_j} r_{hk}\,, X_{ij} = 2 \sum_{(h, k) \in P_i \cap P_j} x_{hk}$, where $P_i$ is the set of lines on the unique path from the substation to bus $i$. Matrix $X$ and $R$ satisfy the following property, which is crucial for the stable control design. 


\begin{proposition}\label{prop:RX_pd}
Suppose $x_{ij}, r_{ij}>0$ for all $(i,j)$. Then, $X$ and $R$ are positive definite matrices. 
\end{proposition}

\subsection{Voltage Control Problem Formulation}
The goal of voltage control is to design control to lead the system voltage to reach the acceptable range $[\underline{v}, \overline{v}]$ under any system operating conditions at the lowest cost. Formally, voltage stability is defined as follows.
\begin{definition}[Voltage stability]
\label{def:voltage_stability}
The closed loop system is stable if for any $\mathbf{v}^{env}$ and $\mathbf{v}(0)$, we have $\mathbf{v}(t)$ converges to the set $S_{v} = \{ \mathbf{v}\in\mathbb{R}^n: \underline{v}_i\leq v_i \leq\bar{v}_i \}$ in the sense that $\lim_{t\rightarrow\infty }\dist(\mathbf{v}(t),S_v) = 0$ and the distance is defined as $\dist(\mathbf{v}(t),S_v) = \min_{\mathbf{v}' \in S_v}||\mathbf{v}(t)-\mathbf{v}'||$.
\end{definition}
Violations of the acceptable voltage range are often caused by a sudden change in the load or the generation, which can damage the appliances of the end users and even cause cascading failures if the system cannot return to the range promptly \cite{haes2019survey}. Therefore voltage stability is a vital requirement for the safe operation of the power systems. 

With the requirement for voltage stability, the optimal voltage control problem can be formulated as follows,
\begin{subequations}
\label{opt:voltage_ctrl}
\begin{align}
\min_{\mathbf{\theta}} \quad & J(\theta)= \int_{t=0}^{\infty} \gamma^t \sum_{i=1}^n c_i({v}_i(t), u_i(t)) dt \label{eq_rl:obj}\\
\text{s.t.}\quad & \mathbf{v}(t) = X\mathbf{q}(t) + \mathbf{v}^{env} \label{eq_rl:dyn}\\
& \dot{q}_i(t) = u_i(t) = g_{\theta_i}(v_i(t))\label{eq_rl:policy}\\
& \text{Voltage stability holds.}\label{eq_rl:constr}
\end{align}
\end{subequations}
The goal of the voltage control problem is to reduce the total cost \eqref{eq_rl:obj}, which consists of two parts: the cost on voltage deviation and the cost of control actions. In particular, we consider $c_i({v}_i(t), u_i(t)) = \eta_1 [\max(v_i(t) - \bar{v}_i,0) + \min(v_i(t) - \underline{v}_i,0)]^2 + \eta_2 (u_i(t))^2$. Here $\eta_1, \eta_2$ are coefficients that balance the cost of action with respect to the voltage deviation.
We can set different $\eta_1, \eta_2$ at different nodes, and for simplicity, we choose the same $\eta_1,\eta_2$ across all nodes in the paper. 
Voltage dynamics of power system are represented by the power flow equation~\eqref{eq_rl:dyn}.  We envision that the reactive power control loop is embedded in an inverter control loop and operate at very fast timescales. Therefore, we use a continuous-time system to model the voltage/reactive power dynamics (unlike the more conventional discrete-time model). The control action $u_i$ means the rate of change of the reactive power injection $q_i$ in equation~\eqref{eq_rl:policy}, and we focus on the class of decentralized polices, $u_i(t) = g_{i,\theta_i}(v_i(t))$ only depends on local voltage measurement $v_i(t)$. Here $\theta_i$ is the policy parameter for the local policy $g_{i,\theta_i}$, and $\theta = (\theta_i)_{i\in\mathcal{N}}$ is the collection of the local policy parameters and is also the decision variable in \eqref{opt:voltage_ctrl}.


\textit{Transient cost vs. stationary cost. } Our problem formulation is different from those in the literature \cite{zhang2013local,bolognani2013distributed,li2014real,zhu2016fast,cavraro2016value,liu2018hybrid,tang2019fast,qu2019optimal,magnusson2020distributed} in the sense that the existing works typically consider the cost in stationarity, meaning the cost is evaluated at the fixed point or stationary point of the system. In contrast, our work considers the transient cost evaluated along the system trajectory, which is also an important metric for the performance of voltage control.
An important future direction is to unify these two perspectives and design policies that can optimize both transient and stationary costs. 


\subsection{Existing Work: DDPG for Optimal Voltage Control}
\label{sec:preliminary_ddpg}
In order to solve the optimal voltage control problem in \eqref{opt:voltage_ctrl}, one need the exact system dynamics, i.e., $X$. However, for distribution system, the exact network parameters are often unknown or hard to estimate in real systems~\cite{chen2020data}. Further, \eqref{opt:voltage_ctrl} is a nonlinear control problem 
Reinforcement learning provides a powerful paradigm for solving \eqref{opt:voltage_ctrl}, by training a policy that maps the state to action via interacting with the environment, so as to minimize the loss function defined as~\eqref{eq_rl:obj}. 

There are many RL algorithms to solve the policy minimization problem \eqref{opt:voltage_ctrl}. In this paper, we focus on the class of RL algorithms called policy optimization. Generally speaking, the procedure is to run gradient methods on the policy parameter $\theta$ with step size $\eta$,
$ \theta \leftarrow \theta - \eta \nabla J(\theta). $
To approximate the gradient $\nabla J(\theta)$, one can use sampled trajectories such as REINFORCE~\cite{williams1992simple} or value function approximation such as actor-critic methods \cite{konda2000actor}. As we are dealing with deterministic policies, one of the most popular choices is the Deep Deterministic Policy Gradient (DDPG)~\cite{lillicrap2015continuous}, where the policy gradient is approximated by
\begin{equation}
\label{eq:policy_update}
\nabla J(\theta) \approx \frac{1}{N} \sum_{i\in B} \nabla_u \hat{Q}_{\phi}(v, u)|_{v = v[i], u=g_{\theta}(v[i])} \nabla_{\theta} g_{\theta}(v)|_{v[i]}\,,
\end{equation}
where $g_{\theta}(v)$ is the actor network, and $\{v[i], u[i]\}_{i\in B}$ are a batch of samples with batch size $|B| = N$ sampled from the replay buffer which stores historical state-action transition tuples. 
Here $\hat{Q}_{\phi}(v, u)$ is the value network (a.k.a critic network) that can be learned via temporal difference learning,
\begin{equation}
    \label{eq:value_update}
    \min_{\phi} L(\phi) = E_{(v, u, r, v')} [Q_{\phi}(v, u) - (r+\gamma Q_{\phi}(v', g_{\theta}(v'))]
\end{equation}
where $v'$ is system voltage after taking action $u$ and realization of $v^{env}$.
For more details of DDPG, readers could refer to~\cite{lillicrap2015continuous}. There have been a growing line of papers that use RL for voltage control~\cite{chen2020data,wang_data-driven_2020,wang_safe_2020}. In particular, ~\cite{wang_data-driven_2020} uses standard multi-agent DDPG for voltage control. However, in standard DDPG, stability is not an explicit requirement. It is more like an implicit regularization, because instability usually leads to high (or infinite) costs. Next, we will introduce our framework that guarantees stability in policy learning.

\section{Main Results}
\label{sec:method}
We now introduce our main framework for stability constrained policy learning for voltage control. We start by stating our main result on the voltage control stability. We demonstrate the voltage stability constraint can be translated to a monotonicity constraint on the policy, which can be satisfied by smart design of monotone neural networks.

\subsection{Key Idea: Monotonicity Guarantees Stability}
As we mentioned in Section~\ref{sec:preliminary_ddpg}, the lack of an explicit stability requirement in standard RL algorithms can lead to several issues. During the training phase, the policy may become unstable, causing the training process to terminate. Even after a policy is trained, there is no formal guarantee that the closed loop system is stable, which hinders the learned policy's deployment in real-world power systems where there is a very strong emphasis on stability. In order to explicitly constrain stability in policy learning, we constrain the search space of policy in a subset of stabilizing controllers from Lyapunov stability theory. Interestingly, we show that monotone policy guarantees stability for voltage control, which is presented in Theorem \ref{thm:voltage_stab}.

\begin{theorem}\label{thm:voltage_stab}
Suppose for all $i$, $g_{i,\theta_i}$ is a continuously differentiable function satisfying $g_{i,\theta_i}(v_i) = 0$ for $v_i \in [\underline{v}_i,\bar{v}_i]$. Further, suppose each $g_{i,\theta_i}$ is strictly monotonically decreasing on $(-\infty,\underline{v}_i]$ and $[\bar{v}_i,\infty)$, and satisfies $\lim_{v_i\rightarrow\infty} |g_{i,\theta_i}(v_i)| = \infty$. Then, the voltage stability defined in Definition~\ref{def:voltage_stability} holds.
\end{theorem}

Theorem \ref{thm:voltage_stab} shows that the voltage stability condition in~\eqref{eq_rl:constr} can be enforced by constraining the policy network to be monotone, which we will introduce in Section~\ref{sec:algorithm_design}.
The key technique that underpins Theorem \ref{thm:voltage_stab} is the Lyapunov stability theory, which involves defining a positive definite function $V(\cdot)$ that decreases along the system trajectory. 
Specially, we use Krasovskii's method~\cite{khalil2002nonlinear} for constructing the Lyapunov function. For the voltage control problem with dynamics $\dot{\mathbf{v}} = f(\mathbf{v}, \mathbf{u}) = X\mathbf{u}$ and $\mathbf{u} = g_\theta(\mathbf{v}) = [ g_{i,\theta_i}(v_i)]_{i\in\mathcal{N}}$, we consider the following Lyapunov function,
\begin{equation}\label{eq:Kra_func}
    V(\mathbf{v}) =\frac{1}{2} f(\mathbf{v}, g_{\theta}(\mathbf{v}))^\top X^{-1} f(\mathbf{v}, g_{\theta}(\mathbf{v}))
\end{equation}
where $X$ is a positive definite matrix by Proposition~\ref{prop:RX_pd}. A sufficient condition for the system to be stable is that, the derivative of the Lyapunov function~\eqref{eq:Kra_func} satisfies the following condition,
\begin{align}
    \frac{d}{dt} V(\mathbf{v}(t)) &=  (\nabla_{\mathbf{v}} {V}(\mathbf{v}))^\top \dot{\mathbf{v}} \nonumber \\
    &= \frac{1}{2} (X\mathbf{u})^\top  \Big[X^{-1} G(\mathbf{v},\theta) + G(\mathbf{v},\theta)^\top X^{-1} \Big] \nonumber\\
    & \quad \quad (X \mathbf{u}) < 0, \forall \mathbf{v} \notin \mathcal{S}_e  \label{eq:lyapunov_decreasing}
\end{align}
where $G(\mathbf{v},\theta) = \frac{\partial f(\mathbf{v}, \mathbf{u})}{\partial \mathbf{v}} + \frac{\partial f(\mathbf{v}, \mathbf{u})}{\partial \mathbf{u}} \frac{\partial \mathbf{u}}{\partial \mathbf{v}} = X\frac{\partial g_{\theta}(\mathbf{v})}{\partial \mathbf{v}}$ is the system Jacobian and $\mathcal{S}_e$ is the set of equilibrium points. This leads to our voltage stability condition, 
\begin{equation}
\label{eq:stability_LMI_Kra}
[\frac{\partial g_{\theta}(\mathbf{v})}{\partial \mathbf{v}} + \frac{\partial g_{\theta}(\mathbf{v})}{\partial \mathbf{v}}^\top \Big] \prec 0\,, \forall \mathbf{v} \notin \mathcal{S}_e 
\end{equation} 
In particular, since the controller is decentralized where $u_i(t) = g_{i,\theta_i}(v_i(t))$ only depends on local voltage measurement $v_i(t)$, the Jacobian matrix $\frac{\partial g(\mathbf{v})}{\partial \mathbf{v}}$ is diagonal with the $i$-th element as $\frac{\partial g_{i,\theta_i}(v_i)}{\partial v_i}$. Therefore, conditions~\eqref{eq:stability_LMI_Kra} can be met with each $g_{i,\theta_i}$ being strictly monotonically decreasing on $(-\infty,\underline{v}_i]$ and $[\bar{v}_i,\infty)$, and $g_{i,\theta_i}(v_i) = 0$ for $v_i \in [\underline{v}_i,\bar{v}_i]$. The detailed proof are as provided in Appendix~\ref{sec:proof}.

\subsection{Algorithm Design}
\label{sec:algorithm_design}
The proposed stability-constrained policy learning algorithm works as follows. At every training iteration $k$, we randomly generate an initial states $\{v_i(0)\}$ for all nodes $i=1,...,N$. Then we use the current policy $g^{(k)}_{i, \theta_i}(v_i(t))$ to generate a trajectory of length $T$, and store the (state, action, reward, next state) data pairs in a replay buffer. Next, we use random samples from the replay buffer to update the policy and value networks following Eq.~\eqref{eq:policy_update} and Eq.~\eqref{eq:value_update}.
Specially, we parameterize the control policies $g_{i,\theta_i}(v_i), \forall i = 1,..., N$ via monotone neural networks with a deadband in $[\underline{v}_i,\bar{v}_i]$. As shown in Theorem \ref{thm:voltage_stab}, such design guarantees voltage stability.

There are different approaches for monotone neural network architecture design in literature~\cite{sill1997monotonic,daniels2010monotone,cui2020reinforcement}.
In this paper we follow the monotonic neural network design in~\cite[Lemma 3]{cui2020reinforcement}, which used a single hidden layer neural network with $d$ hidden units and ReLU activation. 
\begin{corollary}(Stacked ReLU Monotone Network~\cite[Lemma 3]{cui2020reinforcement})
\label{corollary:stacked_relu}
 The stacked ReLU function constructed by Eq~\eqref{eq:relu_pos} is monotonic increasing for $x > 0$ and zero when $x \leq 0$.
\begin{subequations}\label{eq:relu_pos}
    \begin{align}
    \xi^{+}(x; w^+, b^+) &= {(w^+)^\top} \text{ReLU}(\mathbf{1} x + b^+)\\
    \text{where} & \sum_{i=1}^{l} w^{+}_i \geq 0, \forall l = 1, 2, ..., d\\
    & b^{+}_1 = 0, b^{+}_l \leq b^{+}_{l-1}, \forall l =2, 3, ..., d
    \end{align}
\end{subequations}
The stacked ReLU function constructed by Eq~\eqref{eq:relu_neg} is monotonic increasing for $x < 0$ and zero when $x \geq 0$.
\begin{subequations}\label{eq:relu_neg}
    \begin{align}
    \xi^{-}(x; w^{-}, b^{-}) &= (w^{-})^\top \text{ReLU}(-\mathbf{1} x + b^{-})\\
    \text{where} & \sum_{i=1}^{l} w^{-}_i \leq 0, \forall l = 1, 2, ..., d\\
    & b^{-}_1 = 0, b^{-}_l \leq b^{-}_{l-1}, \forall l =2, 3, ..., d
    \end{align}
\end{subequations}
\end{corollary}

Following Corollary~\ref{corollary:stacked_relu}, we parameterize the controller at bus $i$ as $g_{i,\theta_i}(v_i) = -[\xi_{\theta_i}^{+}(v_i) + \xi_{\theta_i}^{-}(v_i)]$ where $\xi_{\theta_i}^{+}(v_i): \mathbb{R} \rightarrow \mathbb{R}$ is monotonically increasing for $v_i >0$ and zero when $v_i \leq 0$, and $\xi_{\theta_i}^{-}(v_i): \mathbb{R} \rightarrow \mathbb{R}$ is monotonically increasing for $v_i <0$ and zero otherwise. In addition, to incorporate the dead-band within range $v_i \in [\underline{v}_i, \overline{v}_i]$, we can simply set $w^{+}_{1} = 0, b^{+}_{2} = -\overline{v}_i$ and $w^{-}_{1} = 0, b^{-}_{2} = -\underline{v}_i$. 
We demonstrate the effectiveness of this approach using a case study in the next section.

\section{Case Study}
\label{sec:experiment}

We end the paper with a case study demonstrating the effectiveness of our approach for stability constrained policy learning for voltage control. 

\begin{figure}[h]
    \centering
    \includegraphics[width=0.48\textwidth,trim=4 4 4 4,clip]{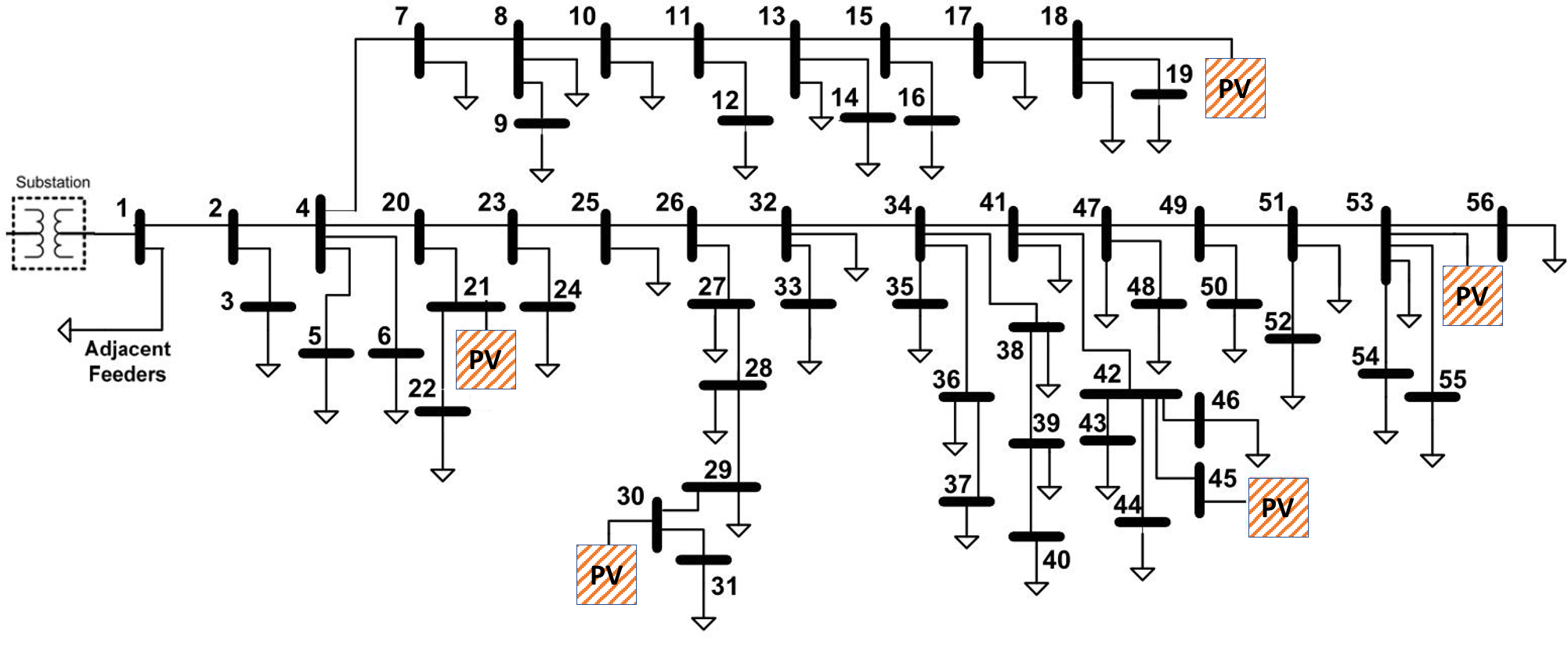}
    \caption{Schematic diagram of SCE 56 bus distribution system with PV generations.}
    \label{fig:SCE56_bus}
\end{figure}

\textbf{Experimental Setup} Our evaluations focus on a Southern California Edison 56 bus distribution system with high penetration of photovoltaic (PV) generations. The detailed system parameters are given in~\cite{farivar2012optimal}. 
Figure \ref{fig:SCE56_bus} provides the 56-bus distribution circuit, where there are 5 PV generators and controllers located at Buses 18, 21, 30, 45 and 53. Simulations of the power system dynamics use pandapower~\cite{pandapower}.

The nominal voltage magnitude at each bus is $12$kV, and the acceptable range for operation is $\pm 5\%$ of the nominal value which is $[11.4\text{kV}, 12.6\text{kV}]$. We simulate three different scenarios: 1) High voltages: the PV generators are generating large amount of power, this corresponds to the day-time scenario in California where there is abundant sunshine that can result in high voltage issues at some buses. 2) Low voltages: the system is serving heavy loads without PV generation. It corresponds to a nighttime scenario when there is no sunshine but significant load, which results in low voltage issues at some buses. 3) A mix of high and low voltages: there are both high PV generation and heavy loads, which results in a mixture of high voltage issues at some buses and low voltages at others. For each scenario, we vary the PV output and the load to obtain different voltage conditions. 

\textbf{Our Approach and Baselines}  We incorporate the monotone policy network design with DDPG~\cite{lillicrap2015continuous} to obtain the provable stable RL controller, and name our approach as Stable-DDPG. For baseline algorithms, we consider the following linear policy with deadband $d_i(v_i) = -[v_i - \overline{v}_i]^{+} + [\underline{v}_i - v_i]^{+}$ (where $[x]^+= \max(x,0)$), which has been proposed and widely used in the power system control community~\cite{li2014real,zhang2013local}. It guarantees stability but may lead to suboptimal control cost. We also compare the performance of Stable-DDPG against DDPG, which is suggested for voltage control in~\cite{wang_data-driven_2020}. Standard DDPG minimizes the control cost without a formal stability guarantee. Details about the implementation and hyperparameters of Stable-DDPG and DDPG are provided in Appendix-\ref{sec:simulation_detail}. 

\begin{figure}[tbp]
    \centering
    \includegraphics[width=0.48\textwidth]{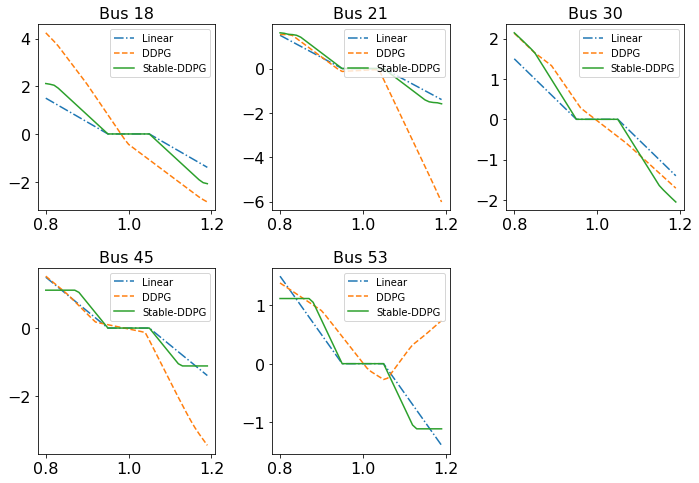}
    \caption{Visualization of DDPG, Stable-DDPG and linear control policy at 5 PV buses. The x-axis is voltage (unit: kV) and the y-axis is control action (unit: MVar).}
    \label{fig:control_policy}
\end{figure}

\textbf{Experimental Results} Figure~\ref{fig:control_policy} illustrates the control policy that is learned from Stable-DDPG and the baseline linear control~\cite{li2014real}.
Standard DDPG does not guarantee stability and thus can lead to ``infinite’’ voltage recovery time and control cost. To obtain a reasonable comparison, we limit the max episode length to be $T= 100$, and compare the voltage recovery time (steps) and reactive power consumption (MVar) on 500 different voltage violation scenarios. Table~\ref{table:voltage_control} shows the results. In terms of control performance, the average voltage recovery time of Stable-DDPG is 31.96 steps, which saves about a third of the response time compared to the linear policy (48.38 steps). 
As the Stable-DDPG is able to drive back the voltage into normal operation state faster, the average transient control cost (computed as the sum of control cost before voltage stabilization) of Stable-DDPG is reduced by $33.38\%$ compared to the linear policy. The standard deviation is high since the recovery time and control effort under different voltage violation conditions can be quite different. 
In addition, the lack of stability guarantee of standard DDPG can lead to higher control and state violation cost, which reflects in Table~\ref{table:voltage_control} that the control performance of standard DDPG is worse than the stable DDPG when averaging all scenarios. 


\begin{table}[t]
\centering
\caption{Performance of linear, DDPG and Stable-DDPG policies on 500 voltage violation scenarios. Note: reactive power consumption denotes the control cost.}
 \label{table:voltage_control}
 \begin{tabular}{lcccc}
    \toprule
    & \multicolumn{2}{c}{Voltage recovery time (steps)}  & \multicolumn{2}{c}{Reactive power $(\text{MVar})$}  \\
    \cmidrule(r){2-5}
    Method     & Mean     & Std & Mean & Std \\
    \midrule
    Linear & 48.38 & 19.57 & 179.08 &   129.03\\
    Stable-DDPG & \textbf{31.96} & 14.49 & \textbf{119.30} & 89.05  \\
    DDPG & 42.87   & 38.32 &  152.53 & 175.62   \\
    \bottomrule
\end{tabular}
\end{table}

\begin{figure}[tbp]
\centering
\begin{minipage}[t]{0.15\textwidth}
  \centering\raisebox{\dimexpr \topskip-\height}{%
  \includegraphics[width=\textwidth]{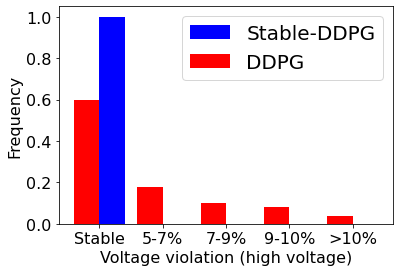}}
  \label{fig:instability}
\end{minipage}
\begin{minipage}[t]{0.15\textwidth}
  \centering\raisebox{\dimexpr \topskip-\height}{%
  \includegraphics[width=\textwidth]{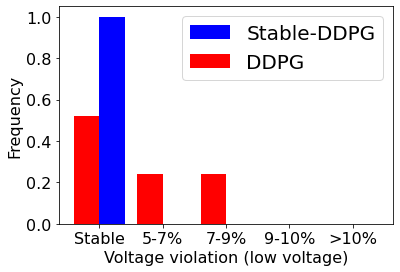}}
  \label{fig:instability}
\end{minipage}
\begin{minipage}[t]{0.15\textwidth}
  \centering\raisebox{\dimexpr \topskip-\height}{%
  \includegraphics[width=\textwidth]{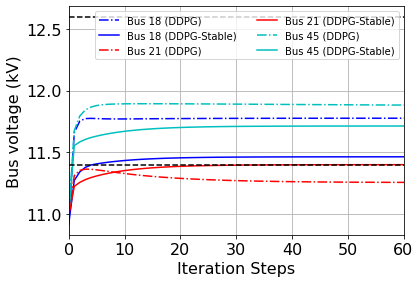}}
  \label{fig:instability}
\end{minipage}
\caption{Voltage stability of DDPG and Stable-DDPG.}
\label{fig:voltage_violation}
\end{figure}

\begin{figure}[t!]
\centering
    \begin{minipage}{0.15\textwidth}
        \centering
        \includegraphics[width=\linewidth]{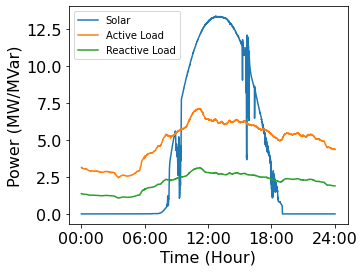}
    \end{minipage}
    \begin{minipage}{0.15\textwidth}
        \centering
        \includegraphics[width=\linewidth]{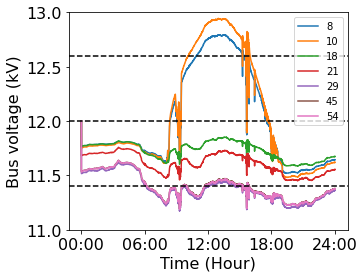}
    \end{minipage}
    \begin{minipage}{0.15\textwidth}
        \centering
        \includegraphics[width=\linewidth]{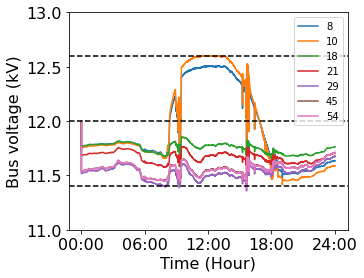}
    \end{minipage}
\caption{Stable-DDPG test with real-world load and PV generation dataset. Left plot is the PV and aggregated load. Right two plots are the voltage without control and with Stable-DDPG.}
\label{fig:real_world}
\end{figure}
Figure \ref{fig:voltage_violation} compares the ability to achieve voltage stability of DDPG and Stable-DDPG under various test scenarios. The left plot shows the histogram of over-voltage ratio (i.e., $(v_T-v_0)^{+}/{v_0}$) and the middle plot shows the under-voltage ratio (i.e., $(v_0-v_T)^{+}/{v_0}$). Our method achieves voltage stability in all scenarios, whereas DDPG may lead to voltage instability, with the final voltage beyond the $\pm 5\%$ range for 30-40\% test scenarios. 
This is a serious issue since large voltage deviations violate regulatory requirements \cite{regulator} and can cause cascading failures \cite{haes2019survey}. 
Figure~\ref{fig:voltage_violation} right shows a test case where DDPG leads to voltage instability at bus 21. 

Finally, we test the proposed method using real-world data from DOE~\cite{qu2019optimal}. We simulate a massive solar penetration scenario where all buses are associated with PV and voltage controllers. 
We compare the voltage dynamics without voltage control and when Stable-DDPG is used. Corresponding to the time resolution of the PV and load trajectory, the proposed method adjust its control output every 6s. The voltage control results are given in Figure~\ref{fig:real_world}. There are severe voltage violations without control, due to the high volatility in load and PV generation. In contrast, Stable-DDPG quickly brings the voltage into the stable operation range, which further demonstrates its applicability in power system voltage control.






\section{Conclusion}
In this work, we propose a stability aware policy learning framework that formally guarantees the stability of RL in safety-critical systems. The key technique that underpins the proposed approach is to use Krasovskii's method for Lyapunov function construction and enforce the stability condition via monotone policy network design. We demonstrate the performance of the proposed method in real-world power system voltage control.  Krasovskii's method is one way to construct the Lyapunov function, and a future direction is to incorporate other principled ways to construct Lyapunov functions in control theory.  

\bibliographystyle{IEEEtran}
\bibliography{reference}

\newpage

\appendix

\section*{Appendix-A: Proof of the Main Results}
\label{sec:proof}
For the proofs of Theorem \ref{thm:voltage_stab}, we use a generalization of Lyapunov's direct method, known as LaSalle's Invariance Principle. We provide a version of it below, adapted from \cite{slotine1991applied}, which we slightly change to fit the rest of the paper. 

\begin{proposition}[Theorem 3.4, 3.5 in \cite{slotine1991applied}]
For dynamical system $\dot{x} = F(x)$, suppose $V:\mathbb{R}^n\rightarrow \mathbb{R}$ is a continuously differentiable function satisfying $V(x)\geq 0$, $\dot{V}(x)= [\nabla_x V(x)]^\top F(x)\leq 0,\forall x\in\mathbb{R}^n$. Let $S_e = \{ x: \dot{V}(x)= 0 \}$. If there exists $a\in\mathbb{R}$ such that the level set $L_a:=\{x: V(x)\leq a\}$ is bounded, then for any $x(0)\in L_a$ we have $\dist(x(t),S_e)\rightarrow 0$. Further, if $V$ is radially unbounded, i.e. $V(x)\rightarrow\infty$ as $\Vert x\Vert\rightarrow\infty$, then, for any $x(0)\in\mathbb{R}^n$, we have $\dist(x(t),S_e)\rightarrow 0$. 
\end{proposition}

With the LaSalle's Invariance Principle, we are now ready to prove Theorem ~\ref{thm:voltage_stab}. 
\begin{proof}[Proof of Theorem \ref{thm:voltage_stab}] For the voltage control problem \eqref{opt:voltage_ctrl}, let $\mathbf{v}(t)$ be the state, $\mathbf{u}(t)$ be the action. Then, we have $\dot{\mathbf{v}} = X \dot{\mathbf{q}} = X \mathbf{u} = X g_\theta(\mathbf{v})$, where $g_\theta(\mathbf{v}) = [ g_{i,\theta_i}(v_i)]_{i\in\mathcal{N}}$ are the decentralized policies. 
We consider the following Lyapunov function,  
$$ V(\mathbf{v}) = \frac{1}{2} g_\theta(\mathbf{v})^\top X g_\theta(\mathbf{v}). $$
Clearly, $V$ is positive definite and is radially unbounded by the assumptions of Theorem~\ref{thm:voltage_stab}. 
By LaSalle's Invariance principle, to prove Theorem~\ref{thm:voltage_stab}, we only need to show the following claim.

\smallskip
\noindent\textbf{Claim:}$\frac{d}{dt}V(\mathbf{v}(t))\leq 0$, and $\frac{d}{dt}V(\mathbf{v}(t))= 0$ only when $\mathbf{v}\in S_v$, where we recall $S_{v} = \{ \mathbf{v}\in\mathbb{R}^n: \underline{v}_i\leq v_i \leq\bar{v}_i \}$. 

It is easy to check that, 
\begin{align}
    \frac{d}{dt} V(\mathbf{v}(t)) &=  (\nabla_{\mathbf{v}} {V}(\mathbf{v}))^\top \dot{\mathbf{v}} \nonumber =  (X g_\theta(\mathbf{v}))^\top   \frac{\partial g_\theta(\mathbf{v})}{\partial \mathbf{v}}   (X g_\theta(\mathbf{v}))  .
\end{align}
Note that $X g_\theta(\mathbf{v}) = 0$ if and only if $\mathbf{v} \in S_v$. Therefore, to show the claim, it suffices to show the following two conditions,
\begin{subequations}\label{eq:stab_kra_weak}
\begin{align}
    \frac{\partial g_\theta(\mathbf{v})}{\partial \mathbf{v}} \preceq 0,&\forall \mathbf{v}\in\mathbb{R}^n, \label{eq:stab_kra_weak_1}\\
    X g_\theta(\mathbf{v})  \not\in \ker(\frac{\partial g_\theta(\mathbf{v})}{\partial \mathbf{v}})/\{0\},& \forall \mathbf{v}\in\mathbb{R}^n,\label{eq:stab_kra_weak_2}
\end{align}
\end{subequations}
where $\ker(\cdot)$ denotes the null space of a matrix. 

We first check~\eqref{eq:stab_kra_weak_1}. Note $\frac{\partial g_\theta(\mathbf{v})}{\partial \mathbf{v}}$ is a diagonal matrix whose $i$'th entry is $g_{i,\theta_i}'(v_i)$, which is nonnegative as  $g_{i,\theta_i}(v_i)$ is monotonically decreasing. Therefore, \eqref{eq:stab_kra_weak_1} is true. To check \eqref{eq:stab_kra_weak_2}, suppose $\mathbf{v}$ is such that,
\begin{align}
    0=  \frac{\partial g_\theta(\mathbf{v})}{\partial \mathbf{v}}  X g_\theta(\mathbf{v}).\label{eq:voltage_stability_condition2}
\end{align}
Then, let $C\subset\mathcal{N}$ be the set of indices $\{i\in\mathcal{N}: v_i\in [\underline{v}_i,\bar{v}_i]\}$, and $C' = \mathcal{N}/C$. In the following, we will show that $C'=\emptyset$. Suppose the contrary is true, i.e. $C' \neq \emptyset$. Note for $i\in C$, $g_{i,\theta_i} (v_i)= 0$, and further, the $i$'th diagonal entry of $\frac{\partial g_\theta(\mathbf{v})}{\partial \mathbf{v}}$ is zero. For $i\in C'$, the $i$'th diagonal entry of $\frac{\partial g_\theta(\mathbf{v})}{\partial \mathbf{v}}$ is non-zero. These show that \eqref{eq:voltage_stability_condition2} is equivalent to 
\begin{align}
    X_{C',C'} g_{C',\theta_{C'}} (\mathbf{v}_{C'}) = 0,\label{eq:appendix:submatrix}
\end{align}
where $X_{C',C'}$ is the submatrix of $X$ corresponding to the rows and columns in $C'$, and $g_{C',\theta_{C'}}(\mathbf{v}_{C'})$ is the subvector of $g_\theta(\mathbf{v})$ corresponding to the indices in $C'$. Since $X_{C',C'}$ is positive definite as it is a principal submatrix of $X$, we have \eqref{eq:appendix:submatrix} indicates that, for any $i\in C'$, $g_{i,\theta_i}(v_i)=0 \Rightarrow v_i\in[\underline{v}_i,\bar{v}_i]$. Per the definition of $C$, this shows that such $i$ must be an element of $C$, which is impossible since $C'$ and $C$ are disjoint. Therefore, we have a contradiction, and as a result, $C'=\emptyset$ and $g_{i,\theta_i}(v_i) = 0, \forall i\in\mathcal{N}$. Therefore, we have $  X g_\theta(\mathbf{v}) = 0$, and hence $X g_\theta(\mathbf{v}) \not\in \ker(\frac{\partial g_\theta(\mathbf{v})}{\partial \mathbf{v}})/\{0\}$. As a result, $\eqref{eq:stab_kra_weak_2}$ holds. Therefore, the claim is proven and the proof of Theorem~\ref{thm:voltage_stab} follows. 
\end{proof}

\section*{Appendix-B: Simulation Details}

We use Pytorch to build all RL models and run the training process in MacBook Pro Personal Laptop with 16 GB 2400 MHz DDR4 memory and 2.2 GHz Intel Core i7 processor. The reward function for training is $c(\mathbf{v}, \mathbf{u}) = 100 (\text{dist}(\mathbf{v}, S_v))^2 + 50 ||\mathbf{u}||_2^2$, where $\mathbf{v}$ is voltage vector and $\mathbf{u}$ is the policy output.
Table~\ref{table:hyperparameter} shows the hyperparameters used for both DDPG and Stable-DDPG. 
For policy network design, Stable-DDPG requires the policy network to be monotone, and the monotone network architecture based on~\cite{cui2020reinforcement} only applies to single layer neural network. Thus, we use a one fully-connected layer neural network for Stable-DDPG and a larger capacity model (two-layer fully-connected neural network) for standard DDPG. As shown in Fig~\ref{fig:training_reward}, Stable-DDPG has higher initial reward and lower variance because of the monotone structure (i.e., stability guarantee by design). 
\label{sec:simulation_detail}
\begin{table}[htbp]
\centering
\caption{Hyperparameters for DDPG and Stable-DDPG}
 \label{table:hyperparameter}
 \renewcommand{\arraystretch}{1.2}
 \begin{tabular}{c|c|c}
    \hline
    Hyper-parameters & DDPG & Stable-DDPG\\
    \hline
    Policy Net & 100-100 & 100\\
    Q function net & 100-100 & 100-100 \\
    Discount factor ($\gamma$)& 0.99 & 0.99\\
    Policy net learning rate & 1e-4& 1e-4\\
    Q function net learning rate & 2e-4 & 2e-4\\
    Action noise & Gaussian(0, 0.05) & Gaussian(0, 0.05)\\
    Maximum replay buffer size & 1000000 & 1000000\\
    Target network update ratio & 1e-2 &1e-2\\
    Batch size &  256 & 256\\
    Activation function & ReLU & ReLU\\
    Training episode & 600 & 200 \\
    Episode length & 30 & 30 \\
    State dimension ($v_i$) & 1 & 1\\
    Action dimension ($u_i$) & 1 & 1\\
    \hline
\end{tabular}
\end{table}
\begin{figure}[htbp]
\centering
\begin{minipage}[t]{0.24\textwidth}
  \centering\raisebox{\dimexpr \topskip-\height}{%
  \includegraphics[width=\textwidth]{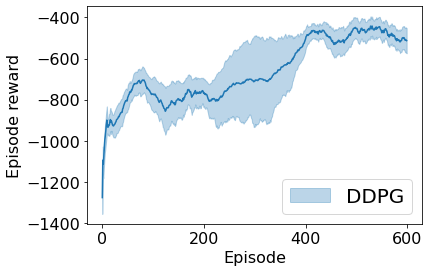}}
  \label{fig:instability}
\end{minipage}
\begin{minipage}[t]{0.24\textwidth}
  \centering\raisebox{\dimexpr \topskip-\height}{%
  \includegraphics[width=\textwidth]{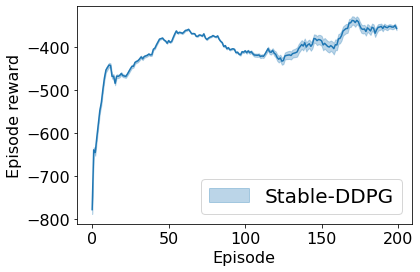}}
  \label{fig:training_reward}
\end{minipage}
\caption{Comparison of training episodic reward from DDPG and Stable-DDPG. The mean and standard deviations are evaluated based on 5 random seeds. }
\label{fig:training_reward}
\end{figure}

\end{document}